\documentclass[aps,preprint,showpacs,superscriptaddress]{revtex4}%
\usepackage{amsfonts}
\usepackage{amsmath}
\usepackage{amssymb}
\usepackage{graphicx}%
\begin{document}
\title{Disentanglement of two qubits coupled to an $XY$ spin chain: Role of quantum
phase transition}
\author{Zi-Gang Yuan}
\affiliation{State Key Laboratory for Superlattices and Microstructures, Institute of
Semiconductors, Chinese Academy of Sciences, P.O. Box 912, Beijing 100083, China}
\author{Ping Zhang}
\affiliation{Institute of Applied Physics and Computational Mathematics, P.O. Box 8009,
Beijing 100088, China}
\author{Shu-Shen Li}
\affiliation{State Key Laboratory for Superlattices and Microstructures, Institute of
Semiconductors, Chinese Academy of Sciences, P.O. Box 912, Beijing 100083, China}

\begin{abstract}
We study the disentanglement of two spin qubits which interact with a general
XY spin-chain environment. The dynamical process of the disentanglement is
numerically and analytically investigated in the vicinity of quantum phase
transition (QPT) of the spin chain in both weak and strong coupling cases. We
find that the disentanglement of the two qubits is in general enhanced greatly
when the environmental spin chain is exposed to QPT. We give a detailed
analysis to facilitate the understanding of the QPT-enhanced decaying behavior
of the disentanglement factor. Furthermore, the scaling behavior in the
disentanglement dynamics is also revealed and analyzed.

\end{abstract}
\maketitle

The coupling between an entangled quantum system and its environment leads to
disentanglement of the system, the process through which quantum information
is degraded. Disentanglement is a crucial issue that is of fundamental
interest due to the fact that the distributed nonlocal coherence among
multi-particles by the entanglement really matters in many important
applications of quantum information \cite{Preskill,Nielson}. Consequently, the
fragility of nonlocal entanglement is recognized as a main obstacle to
realizing quantum computing and quantum information processing (QIP)
\cite{Viola1999,Beige2001}. Apart from the important link to QIP realizations,
a deeper understanding of disentanglement is also expected to lead new
insights into quantum fundamentals, particularly quantum measurement and
quantum-classical transitions \cite{Dodd}. Recently, Yu and Eberly
\cite{Yu2004} have showed that two entangled qubits become completely
disentangled in a finite time under the influence of pure vacuum noise.
Zubairy et al. \cite{Zubairy} have demonstrated how the high quality cavities
can be used to realize the new class of quantum erasers referred to as quantum
disentanglement erasers. Dodd \cite{Dodd} has studied the competing effects of
environmental noise and interparticle coupling on disentanglement by solving
the dynamics of two harmonically coupled oscillators. Cucchietti et al.
\cite{Cucchietti2005} have considered the decoherence effect of a
non-interacting spin chain on a single qubit.

In this paper, we study the disentanglement dynamics of a two-qubit quantum
sysmtem. Here, the key point is that we choose a special correlated $XY$ spin
chain to model the surrounding environment. This choice of the correlated
environment is directly motivated by the recent recognition that the
single-qubit decoherence induced by a spin-chain environment displays highly
interesting properties \cite{Quan,Cucchietti2007,Yuan,Yi} due to the unique
occurrence of quantum phase transition (QPT) in the spin-chain environmental
subsystem. Quan et al. \cite{Quan} have studied the transition dynamics of a
quantum two-level system from a pure state to a mixed one induced by QPT of
the surrounding many-body system. They have shown that the decaying behavior
of the Loschmidt echo (LE) is best enhanced by QPT of the surrounding system.
Cucchietti et al. \cite{Cucchietti2007} have found that the QPT of the
spin-chain environment will drive the decay of the quantum coherences in the
central quantum system to be Gaussian with a width independent of the
system-environment coupling strength.

Motivated by the above-mentioned advances in the QPT effect on the
single-qubit decoherence, we turn to study the QPT effect of the environmental
spin chain on the two-qubit disentanglement of the central quantum system. The
coupled spin system we consider in this paper consists of two quantum
subsystems. One subsystem is characterized by two spin-1/2 Hamiltonians, which
denotes the general two qubits. We call this subsystem the central system, in
the sense that these two spins play the role of measuring disentanglement.
Whereas the other subsystem (a general $XY$ spin chain in a transverse
magnetic field) plays the role of the many-body environment. Compared to the
Ising model which has been recently used to study the QPT effect on the
disentanglement \cite{Sun2007}, the XY model is parameterized by $\gamma$ and
$\lambda$ (see Eq. (\ref{1}) below). Two distinct critical regions appear in
parameter space: the segment $(\gamma,\lambda)=(0,(0,1))$ for the $XX$ spin
chain and the critical line $\lambda_{c}=1$ for the whole family of the $XY$
model \cite{Sach}.

The total Hamiltonian for two central spins transversely coupled to a
environmental spin chain, which is described by the one-dimensional $XY$
model, is given by ($\hbar$ is taken to be unity)
\begin{align}
H  &  =-\sum_{l}^{N}\left(  \frac{1+\gamma}{2}\sigma_{l}^{x}\sigma_{l+1}%
^{x}+\frac{1-\gamma}{2}\sigma_{l}^{y}\sigma_{l+1}^{y}+\lambda\sigma_{l}%
^{z}\right) \label{1}\\
&  -\frac{g}{2}\left(  \sigma_{A}^{z}+\sigma_{B}^{z}\right)  \sum_{l}%
^{N}\sigma_{l}^{z}\nonumber\\
&  \equiv H_{E}^{(\lambda)}+H_{I}.\nonumber
\end{align}
Where $H_{E}^{(\lambda)}$ given by first line in Eq. (\ref{1}) denotes the
Hamiltonian of the environmental spin chain, and $H_{I}$ given by the second
line denotes describes the interaction between the central two-qubit spins and
the spin chain. The Pauli matrices $\sigma_{A(B)}^{z}$ ($\alpha$=$x,y,z$) and
$\sigma_{l}^{\alpha}$ are used to describe the central two-qubit spins and the
environmental spin-chain subsystems, respectively. The parameters $\lambda$
characterizes the intensity of the transverse magnetic field, and $\gamma$
measures the anisotropy in the in-plane interaction. It is well known that the
$XY$ spin model described by the first line in Eq. (\ref{1}) encompasses two
other well-known spin models: the Ising spin chain with $\gamma$=$1$ and the
$XX$ chain with $\gamma$=$0$.

The eigenstates of the operator $(\sigma_{A}^{z}+\sigma_{B}^{z})$ are simply
given by%
\begin{align}
|1\rangle &  =|++\rangle_{AB}\text{, \ }|2\rangle=|--\rangle_{AB},\label{2}\\
|3\rangle &  =\frac{1}{\sqrt{2}}(|+-\rangle_{AB}+|+-\rangle_{AB}),\nonumber\\
|4\rangle &  =\frac{1}{\sqrt{2}}(|+-\rangle_{AB}-|+-\rangle_{AB}),\nonumber
\end{align}
where $|\pm\pm\rangle_{AB}\equiv|\pm\rangle_{A}\otimes|\pm\rangle_{B}$ denote
the eigenstates of the product Pauli spin operator $\sigma_{A}^{z}%
\otimes\sigma_{B}^{z}$ with eigenvalues $\pm1$. The two-qubit states
$|1\rangle$, $|2\rangle$, and $|3\rangle$ are simply spin triplet states with
total central spin $\sigma_{AB}=2$, while $|4\rangle$ is singlet state with
total central spin $\sigma_{AB}=0$. In terms of these two-spin states, the
Hamiltonian (\ref{1}) is rewritten as
\begin{equation}
H=\sum_{j=1}^{4}|j\rangle\langle j|\otimes H_{E}^{(\lambda_{j})}, \label{3}%
\end{equation}
where the parameters $\lambda_{j}$ are
\begin{equation}
\lambda_{1(2)}=\lambda\pm g\text{, }\lambda_{3}=\lambda_{4}=\lambda, \label{4}%
\end{equation}
and $H_{E}^{(\lambda_{j})}$ is given from $H_{E}^{(\lambda)}$ by the
replacement of $\lambda$ with $\lambda_{j}$.

As for quantum criticality in the $XY$ model, there are two universality
classes depending on the anisotropy $\gamma$. The critical features are
characterized by a critical exponent $\nu$ defined by $\xi\sim|\lambda
-\lambda_{c}|^{-\nu}$ with $\xi$ representing the correlation length. For any
value of $\gamma$, quantum criticality occurs at a critical magnetic field
$\lambda_{c}$=$1$. For the interval $0<\gamma\leq1$ the model belongs to the
Ising universality class characterized by the critical exponent $\nu$=$1$,
while for $\gamma$=$0$ the model belongs to the $XX$ universality class with
$\nu$=$1/2$ \cite{Sach}.

Considering the initial state $|\Psi(0)\rangle$=$|\phi_{S}(0)\rangle
\otimes|\psi_{E}(0)\rangle$, where $|\phi_{S}(0)\rangle$ is the initial state
for the two central spins and $|\psi_{E}(0)\rangle$ is the initial state for
the environmental spin chain, then the subsequent time evolution of the
coupled spin system is determined by the time evolution operator $U(t)$%
=$\exp(-iHt)$, $|\Psi(t)\rangle$=$U(t)|\Psi(0)\rangle$. Given $|\Psi
(t)\rangle$, the central quantity for our investigation, i.e., the evolved
reduced density matrix for the two central spins, will be straightforward to
obtain. Thus the key task is to determine the time evolution operator in a
maximally compact form. For this purpose, we follow the standard procedure
\cite{Sach} by defining the conventional Jordan-Wigner (JW) transformation
\begin{align}
\sigma_{l}^{x}  &  =\underset{m<l}{\prod}(1-2a_{m}^{+}a_{m})\left(
a_{l}+a_{l}^{+}\right)  ,\nonumber\\
\sigma_{l}^{y}  &  =-i\underset{m<l}{\prod}(1-2a_{m}^{+}a_{m})\left(
a_{l}-a_{l}^{+}\right)  ,\label{e2}\\
\sigma_{l}^{z}  &  =1-2a_{l}^{+}a_{l},\nonumber
\end{align}
which maps spins to one-dimensional spinless fermions with creation
(annihilation) operators $a_{l}^{+}$ ($a_{l}$). After a straightforward
derivation, the projected environmental Hamiltonian becomes%
\begin{equation}
H_{E}^{(\lambda_{j})}=-\overset{N}{\underset{l}{\sum}}[(a_{l+1}^{+}a_{l}%
+a_{l}^{+}a_{l+1})+\gamma(a_{l+1}a_{l}+a_{l}^{+}a_{l+1}^{+})+\lambda
_{j}(1-2a_{l}^{+}a_{l})]. \label{e3}%
\end{equation}
Next we introduce Fourier transforms of the fermionic operators described by
$d_{k}$=$\frac{1}{\sqrt{N}}\sum_{l}a_{l}e^{-i2\pi lk/N}$ with $k$=$-M,..,M$
and $M$=($N-1)/2$. The Hamiltonian (\ref{1}) can be diagonalized by
transforming the fermion operators to momentum space and then using the
Bogoliubov transformation. The final result is%
\begin{equation}
H_{E}^{(\lambda_{j})}=\sum_{k}\Omega_{k}^{(\lambda_{j})}(b_{k,\lambda_{j}}%
^{+}b_{k,\lambda_{j}}-\frac{1}{2}), \label{7}%
\end{equation}
where the energy spectrum $\Omega_{k}^{(\lambda_{j})}$ is given by%
\begin{equation}
\Omega_{k}^{(\lambda_{j})}=2\sqrt{\left(  \epsilon_{k}^{(\lambda_{j})}\right)
^{2}+\gamma^{2}\sin^{2}\frac{2\pi k}{N}} \label{8}%
\end{equation}
with $\epsilon_{k}^{(\lambda_{j})}$=$\lambda_{j}-\cos\frac{2\pi k}{N}$, and
the corresponding Bogoliubov-transformed fermion operators are defined by%
\begin{equation}
b_{k,\lambda_{j}}=\cos\frac{\theta_{k}^{(\lambda_{j})}}{2}d_{k}-i\sin
\frac{\theta_{k}^{(\lambda_{j})}}{2}d_{-k}^{+} \label{9}%
\end{equation}
with angles $\theta_{k}^{(\lambda_{j})}$ satisfying $\cos\theta_{k}%
^{(\lambda_{j})}$=$2\epsilon_{k}^{(\lambda_{j})}/\Omega_{k}^{(\lambda_{j})}$.
It is straightforward to see that the normal mode $b_{k,\lambda_{j}}$ dressed
by the system-environment interaction is related to the purely environmental
normal mode $b_{k,\lambda}$ by the following identity%
\begin{equation}
b_{k,\lambda_{j}}=(\cos\alpha_{k}^{(\lambda_{j})})b_{k,\lambda}-i(\sin
\alpha_{k}^{(\lambda_{j})})b_{-k,\lambda}^{+}, \label{10}%
\end{equation}
where $\alpha_{k}^{(\lambda_{j})}$=$(\theta_{k}^{(\lambda_{j})}-\theta
_{k}^{(\lambda)})/2$.

The time evolution operator for the Hamiltonian (\ref{3}) is then given by%
\begin{equation}
U(t)=\sum_{j=1}^{4}|j\rangle\langle j|\otimes U_{E}^{(\lambda_{j})}(t),
\label{11}%
\end{equation}
where $U_{E}^{(\lambda_{j})}(t)$=$\exp(-iH_{E}^{(\lambda_{j})}t)$ is the
projected time evolution operator for the spin chain dressed by the
system-environment interaction parameter $\lambda_{j}$.

Suppose that initially the central spins $A$ and $B$ are entangled with each
other but not with the spin chain, i.e., at $t$=$0$ the two central spins and
the environmental spin chain are assumed to be described by the product state%
\begin{equation}
|\Psi_{\text{tot}}(0)\rangle=|\phi\rangle_{AB}\otimes|\psi_{E}\rangle,
\label{12}%
\end{equation}
where $|\phi\rangle_{AB}$ is the entangled initial state of the two central
spins and $|\psi_{E}\rangle$ is the initial state of the environmental spin
chain. The evolved reduced density matrix of the central spins is derived to
be%
\begin{align}
\rho_{AB}(t)  &  =\text{Tr}_{E}|\Psi_{\text{tot}}(t)\rangle\langle
\Psi_{\text{tot}}(t)|\label{13}\\
&  =\sum_{j,j^{\prime}=1}^{4}c_{j}c_{j^{\prime}}^{\ast}\langle\psi_{E}%
|U_{E}^{+(\lambda_{j^{\prime}})}(t)U_{E}^{(\lambda_{j})}(t)|\psi_{E}%
\rangle|j\rangle\langle j^{\prime}|,\nonumber
\end{align}
where $c_{j}$=$\langle j|\phi\rangle_{AB}$. Equation (\ref{13}) is our
starting point for the following derivation and discussions. It reveals in Eq.
(\ref{13}) that the environmental spin chain only modulates the off-diagonal
terms of $\rho_{AB}$ through the \textquotedblleft decoherence
factor\textquotedblright%
\begin{equation}
F(t)=\langle\psi_{E}|U_{E}^{+(\lambda_{j^{\prime}})}(t)U_{E}^{(\lambda_{j}%
)}(t)|\psi_{E}\rangle. \label{14}%
\end{equation}
Whereas, the diagonal terms of $\rho_{AB}$ are not influenced by the
environment since for $j$=$j^{\prime}$, the decoherence factor remains unity.
One can see from Eq. (\ref{14}) that the decoherence factor reflects the
overlap between the two states of the environment obtained by evolving the
initial state $|\psi_{E}\rangle$ with two Hamiltonians $H_{E}^{(\lambda_{j})}$
and $H_{E}^{(\lambda_{j^{\prime}})}$, which are different (for $j\neq
j^{\prime}$) by the system-dependent parameters $\lambda_{j}$ and
$\lambda_{j^{\prime}}$ [see Eq. (\ref{7})]. Furthermore, we notice that
similar to the single-qubit case, the present decoherence factor $F(t)$ of the
two qubits also in some special cases has a form of the Loschmidt echo (or
fidelity), which can show universal behavior (with exponential decay) when
$H_{E}^{(\lambda_{j})}$ are classically chaotic Hamiltonians
\cite{Jalabert2001,Gorin2006}. The new physical connotation endowed by the
special choice of spin-chain environment is QPT, which due to its dynamic
hypersensitivity to the perturbation induced by a single qubit as previously
investigated \cite{Quan,Cucchietti2007,Yuan,Yi}, or two qubits to be studied
here, will play a fundamental role in determining the dynamics of the central
spin(s) and the corresponding decoherence (disentanglement) behaviors.

Before proceeding the discussion, we would like to point out that the reduced
density matrix $\rho_{AB}$ sensitively depends through $F(t)$ on the special
choice of the initial central-spin state $|\phi\rangle_{AB}$ and spin-chain
state $|\psi_{E}\rangle$. In particular, if $|\phi\rangle_{AB}$ lies in the
subspace spanned by $|3\rangle$ and $|4\rangle$ [see Eq. (\ref{2})], then
there is no dynamic correlation between central spins and spin-chain
environment, i.e., $F(t)$=$1$ in this case. Thus we choose the initial state
of the central spins to have an entangled form%
\begin{align}
|\phi\rangle_{AB}  &  =a|1\rangle+b|2\rangle\label{15}\\
&  =a|++\rangle_{AB}+b|--\rangle_{AB}.\nonumber
\end{align}
As a consequence, the time evolution of the two central spins will be confined
within this two-dimensional subspace consisting of $|1\rangle$ and $|2\rangle$
and $\rho_{AB}$ is reduced to a $2\times2$ matrix. On the other side, the
choice of the initial spin-chain state $|\psi_{E}\rangle$ also needs to be
mentioned. In the previous work \cite{Quan,Cucchietti2007,Yuan,Yi} involving
decoherence of single qubit in the spin-chain environment, the qubit is chosen
to initially be its unperturbed ground state $|g\rangle$. Then $|\psi
_{E}\rangle$ is naturally and simply chosen to be the ground state of the
constrained spin-chain Hamiltonian, $H_{g}$=$\langle g|H|g\rangle$. In the
present two-qubit case, however, since the initially chosen entangled state
$|\phi\rangle_{AB}$ is not the eigenstate of the unperturbed qubits, thus one
cannot choose the initial state of the spin chain in the same way as used in
the single-qubit discussions. Here we choose the initial state $|\psi
_{E}\rangle$ of the environment to be the ground state $|G\rangle_{\lambda}$
of the purely spin-chain Hamiltonian $H_{E}^{(\lambda)}$. This choice of
$|\psi_{E}\rangle$ is natural since it may be assumed that the coupling
between the central spin subsystem and the spin-chain subsystem is
adiabatically applied.

The ground state $|G\rangle_{\lambda}$ of $H_{E}^{(\lambda)}$ is the vacuum of
the fermionic modes described by $b_{k,\lambda}|G\rangle_{\lambda}$=$0$, and
can be written as $|G\rangle_{\lambda}$=$\prod\nolimits_{k=1}^{M}\left(
\cos\frac{\theta_{k}^{(\lambda)}}{2}|0\rangle_{k}|0\rangle_{-k}+i\sin
\frac{\theta_{k}^{(\lambda)}}{2}|1\rangle_{k}|1\rangle_{-k}\right)  $, where
$|0\rangle_{k}$ and $|1\rangle_{k}$ denote the vacuum and single excitation of
the $k$th mode $d_{k}$, respectively. Note that the ground state is a tensor
product of states, each lying in the two-dimensional Hilbert space spanned by
$|0\rangle_{k}|0\rangle_{-k}$ and $|1\rangle_{k}|1\rangle_{-k}$. From the
relationship between the Bogoliubov modes $b_{k,\lambda}$ and $b_{k,\lambda
_{j}}$ [equation (\ref{10})], one can see that the ground state $|G\rangle
_{\lambda}$ of the purely spin-chain Hamiltonian $H_{E}^{\lambda}$ can be
obtained from the ground state $|G\rangle_{\lambda_{j}}$ of the qubit-dressed
Hamiltonian $H_{E}^{\lambda_{j}}$ by the transformation
\begin{equation}
|G\rangle_{\lambda}=\prod\nolimits_{k=1}^{M}(\cos\alpha_{k}^{(\lambda_{j}%
)}+i\sin\alpha_{k}^{(\lambda_{j})}b_{k,\lambda_{j}}^{+}b_{-k,\lambda_{j}}%
^{+})|G\rangle_{\lambda_{j}}. \label{16}%
\end{equation}

Given the initial state $|\Psi_{\text{tot}}(0)\rangle$=$|\phi\rangle
_{AB}\otimes|\psi_{E}\rangle$ of the whole system, then our present task is to
derive the explicit expression for the decoherence factor $F(t)$. First one
notices that $F(t)$ in Eq. (\ref{14}) can be written as
\begin{align}
|F(t)|  &  =|_{\lambda}\langle G|U_{E}^{+(\lambda_{2})}(t)U_{E}^{(\lambda
_{1})}(t)|G\rangle_{\lambda}|\label{17}\\
&  =|_{\lambda_{2}}\langle G|\prod_{k}\left(  \cos\alpha_{k}^{(\lambda_{2}%
)}-i\sin\alpha_{k}^{(\lambda_{2})}b_{-k,\lambda_{2}}b_{k,\lambda_{2}}\right)
\nonumber\\
&  \times e^{iH_{E}^{(\lambda_{2})}t}e^{-iH_{E}^{(\lambda_{1})}t}\prod
_{k}\left(  \cos\alpha_{k}^{(\lambda_{1})}+i\sin\alpha_{k}^{(\lambda_{1}%
)}b_{k,\lambda_{j}}^{+}b_{-k,\lambda_{1}}^{+}\right)  |G\rangle_{\lambda_{1}%
}|.\nonumber
\end{align}
By using the identity $e^{-iH_{\lambda}t}b_{k,\lambda}^{+}e^{iH_{\lambda}t}%
$=$b_{k,\lambda}^{+}e^{-i\Omega_{k}^{(\lambda)}t}$, Eq. (17) is rewritten as
\begin{align}
\left\vert F\left(  t\right)  \right\vert  &  =\prod\nolimits_{k>0}|\sin
\alpha_{k}^{(\lambda_{1})}\sin\alpha_{k}^{(\lambda_{2})}\cos\left(  \alpha
_{k}^{(\lambda_{1})}-\alpha_{k}^{(\lambda_{2})}\right)  \exp\left(
-i\Omega_{k}^{\lambda_{1}}t+i\Omega_{k}^{\lambda_{2}}t\right) \label{17.5}\\
&  -\cos\alpha_{k}^{(\lambda_{1})}\sin\alpha_{k}^{(\lambda_{2})}\sin\left(
\alpha_{k}^{(\lambda_{1})}-\alpha_{k}^{(\lambda_{2})}\right)  \exp\left(
i\Omega_{k}^{\lambda_{1}}t+i\Omega_{k}^{\lambda_{2}}t\right) \nonumber\\
&  +\sin\alpha_{k}^{(\lambda_{1})}\cos\alpha_{k}^{(\lambda_{2})}\sin\left(
\alpha_{k}^{(\lambda_{1})}-\alpha_{k}^{(\lambda_{2})}\right)  \exp\left(
-i\Omega_{k}^{\lambda_{1}}t-i\Omega_{k}^{\lambda_{2}}t\right) \nonumber\\
&  +\cos\alpha_{k}^{(\lambda_{1})}\cos\alpha_{k}^{(\lambda_{2})}\cos\left(
\alpha_{k}^{(\lambda_{1})}-\alpha_{k}^{(\lambda_{2})}\right)  \exp\left(
i\Omega_{k}^{\lambda_{1}}t-i\Omega_{k}^{\lambda_{2}}t\right)  |.\nonumber
\end{align}
Equation (\ref{17.5}) will be used in the latter discussions, one variant of
its form, which will also be used for discussion, is the following

\begin{align}
|F(t)|  &  =\prod\nolimits_{k>0}\{1-\sin^{2}\left(  2\alpha_{k}^{(\lambda
_{1})}\right)  \sin^{2}\left(  \Omega_{k}^{(\lambda_{1})}t\right)  -\sin
^{2}\left(  2\alpha_{k}^{(\lambda_{2})}\right)  \sin^{2}\left(  \Omega
_{k}^{(\lambda_{2})}t\right) \label{18}\\
&  +2\sin\left(  2\alpha_{k}^{(\lambda_{1})}\right)  \sin\left(  2\alpha
_{k}^{(\lambda_{2})}\right)  \sin\left(  \Omega_{k}^{(\lambda_{1})}t\right)
\sin\left(  \Omega_{k}^{(\lambda_{2})}t\right)  \cos\left(  \Omega
_{k}^{(\lambda_{1})}t-\Omega_{k}^{(\lambda_{2})}t\right) \nonumber\\
&  -4\sin\left(  2\alpha_{k}^{(\lambda_{1})}\right)  \sin\left(  2\alpha
_{k}^{(\lambda_{2})}\right)  \sin^{2}\left(  \alpha_{k}^{(\lambda_{1})}%
-\alpha_{k}^{(\lambda_{2})}\right)  \sin^{2}\left(  \Omega_{k}^{(\lambda_{1}%
)}t\right)  \sin^{2}\left(  \Omega_{k}^{(\lambda_{2})}t\right)  \}^{\frac
{1}{2}}\nonumber\\
&  \equiv\prod\nolimits_{k>0}F_{k}(t).\nonumber
\end{align}

Equation (\ref{18}) [or Eq. (\ref{17.5})] is one main result in this paper. It
can be simplified under some special conditions. For example, if one chooses
the initial spin-chain state to be $|\psi_{E}\rangle$=$|G\rangle_{\lambda_{2}%
}$, then Eq. (\ref{14}) and corresponding Eq. (\ref{18}) will be reduced to a
LE form given in Ref. \cite{Yuan}. It is straightforward to see that each
factor $F_{k}$ in Eq. (\ref{18}) has a norm less than unity, thus one may
expect $F(t)$ to decrease to zero in the large $N$ limit under some reasonable
conditions. Now we study in detail the critical behavior of the decoherence
factor $F(t)$ near the critical point $\lambda_{c}$=$1$ for finite lattice
size $N$ of the spin chain. Following Ref. \cite{Quan}, let us first make a
heuristic analysis of the features of $F(t)$. For a cutoff frequency $K_{c}$
we define the partial product for $F(t)$
\begin{equation}
|F_{c}(t)|=\prod\nolimits_{k=1}^{K_{c}}F_{k}\geq|F(t)|,\label{19}%
\end{equation}
and the corresponding partial sum $S(t)$=$\ln|F_{c}(t)|\equiv-\sum
\nolimits_{k=1}^{K_{c}}|\ln F_{k}|$. For small $k$ and small $g$ (weak
coupling) one has%
\begin{align}
\Omega_{k}^{(\lambda)} &  \approx2\left\vert \lambda-1\right\vert
+O(k^{2}),\label{20}\\
\Omega_{k}^{(\lambda_{j})} &  \approx2\left\vert \lambda_{j}-1\right\vert
+O(k^{2}),\nonumber
\end{align}
and then
\begin{align}
\sin\left(  2\alpha_{k}^{(\lambda_{j})}\right)   &  \approx\frac{\mp2\gamma\pi
kg}{N|(\lambda_{j}-1)(\lambda-1)|},\label{21}\\
\sin\left(  \alpha_{k}^{(\lambda_{1})}-\alpha_{k}^{(\lambda_{2})}\right)   &
\approx\frac{-2\gamma\pi kg}{N|(\lambda_{1}-1)(\lambda_{2}-1)|}.\nonumber
\end{align}
As a result, one has
\begin{align}
S(t) &  \approx-\frac{1}{2}E\left(  K_{c}\right)  \gamma^{2}g^{2}\left(
\lambda-1\right)  ^{-2}\left(  \lambda_{1}-1\right)  ^{-2}\label{22}\\
&  \times\left(  \lambda_{2}-1\right)  ^{-2}\{\left(  \lambda_{2}-1\right)
^{2}\sin^{2}\left(  2\left\vert \lambda_{1}-1\right\vert t\right)  \nonumber\\
&  +\left(  \lambda_{1}-1\right)  ^{2}\sin^{2}\left(  2\left\vert \lambda
_{2}-1\right\vert t\right)  \nonumber\\
&  -2\left\vert \left(  \lambda_{1}-1\right)  \left(  \lambda_{2}-1\right)
\right\vert \sin\left(  2\left\vert \lambda_{1}-1\right\vert t\right)
\nonumber\\
&  \times\sin\left(  2\left\vert \lambda_{2}-1\right\vert t\right)
\cos\left(  4\lambda t\right)  \},\nonumber
\end{align}
where $E\left(  K_{c}\right)  =4\pi^{2}K_{c}\left(  K_{c}+1\right)  \left(
2K_{c}+1\right)  /\left(  6N^{2}\right)  $. In the derivation of the above
equation, we have omitted the terms related to the sum of $k^{4}/N^{4}$.
Consequently, in the short time $t$ one has%

\begin{equation}
\left\vert F_{c}(t)\right\vert \approx e^{-\tau t^{2}} \label{23}%
\end{equation}
when $\lambda\rightarrow\lambda_{c}=1$, where $\tau=8E\left(  K_{c}\right)
\gamma^{2}g^{2}/\left(  \lambda-1\right)  ^{2}$.

One can see from Eq. (\ref{23}) that when $N$ is large enough and $\lambda$
$\rightarrow\lambda_{c}=1$, then $\left\vert F_{c}(t)\right\vert $ will decay
to zero in a short time. It should be noticed that when increasing $N$, the
cutoff frequency $K_{c}$ should also linearly increase to remain the validity
of Eq. (\ref{23}). Otherwise, one would derive an unphysical conclusion that
in the thermodynamic limit, i.e., the number $N$ of the sites approaching
infinite while keeping the length of the spin chain fixed, $\tau$ tends to
zero and thus the approximate expression $\left\vert F_{c}(t)\right\vert $
remains unity without any decay. Therefore, in using Eq. (\ref{23}) to reveal
the close relationship between the decaying behavior of $\left\vert
F(t)\right\vert $ and QPT which occur only in the thermodynamic limit, it is
necessary to keep the value of $K_{c}/N$ invariant when increasing $N$ to
infinity. Such kind of scaling relation will be further revealed in the latter
discussions in this paper. %

\begin{figure}[tbp]
\begin{center}
\includegraphics[width=0.7\linewidth]{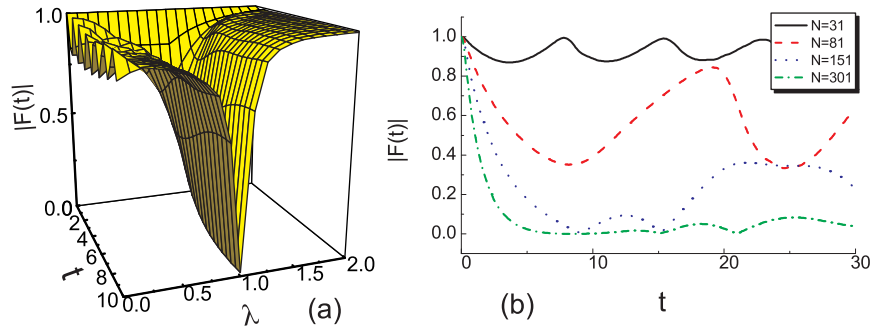}
\end{center}
\caption
{(Color online). (a) Disentanglement factor $|F(t)|$ as a function of magnetic intensity
$\lambda$ and time $t$ for two central spin qubits coupled (with coupling
strength $g=0.05$) to an Ising ($\gamma=1.0$) spin chain with the size
$N=201$. (b) Disentanglement factor for different sizes of Ising spin chain
at QPT point ($\lambda=1$).}
\label{fig1}
\end{figure}%
Now we check the dynamical property of $\left\vert F(t)\right\vert $ by
numerical analysis calculated from the exact expression Eq. (\ref{18}). In
Fig. 1(a), the $\left\vert F(t)\right\vert $ is plotted as a function of
magnetic intensity $\lambda$ and time $t$ for $N$=$201$, $g$=$0.05$, and
$\gamma$=$1.0$ (i.e., the case of Ising spin chain and in the weak coupling
regime). One can see that apart from the critical point $\lambda_{c}$, the
$\left\vert F(t)\right\vert $ in time domain is characterized by an
oscillatory localization behavior. When the amplitude of $\lambda$ approaches
to $\lambda_{c}$, the degree of localization of $\left\vert F(t)\right\vert $
is decreased to zero. The fundamental change occurs at a critical point of
QPT, ie., $\lambda$=$\lambda_{c}$=1. At this point, as revealed in Fig. 1(a),
the $\left\vert F(t)\right\vert $ evolves from unity to zero in a very short
time, which implies that the disentanglement of two central spins is best
enhanced by QPT in the environmental spin chain. The size dependence of the
decoherence factor is shown in Fig. 1(b) for $\lambda$=$\lambda_{c}$ and
$g$=$0.05$. Not surprisingly, with increasing $N$ towards thermodynamic limit,
the role of QPT in Ising spin chain becomes clear by completely disentangling
the two central qubits in a very short time.%

\begin{figure}[tbp]
\begin{center}
\includegraphics[width=0.6\linewidth]{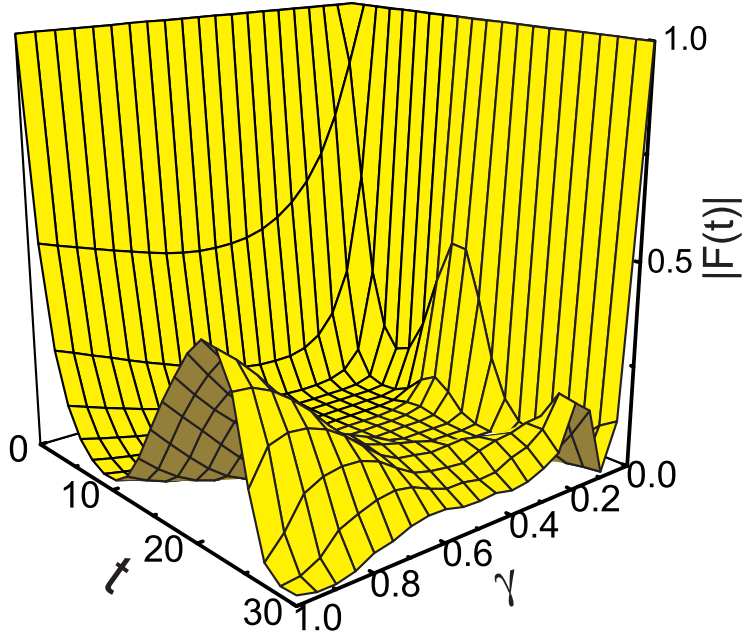}
\end{center}
\caption
{(Color online). Disentanglement factor $|F(t)|$ as a function of spin anisotropy
parameter $\gamma$ and time $t$ for two central spin qubits coupled to an XY
spin chain. The other parameters are set to be $\lambda=1$, $N=201$, and
$g=0.05$.}
\label{fig2}
\end{figure}%
As mentioned at the beginning of this paper, for the XY model we employed,
there are two distinct critical regions in parameter space. Region I is the
segment $(\gamma,\lambda)=(0,(0,1))$ for the $XX$ spin chain, while region II
is a critical line $\lambda_{c}$=$1$ for the whole family of the $XY$ model
(including the special case $\gamma$=$1$ of Ising model). We find that the
best-enhancement behavior of the disentanglement factor $\left\vert
F(t)\right\vert $ only occurs in QPT region II except for the point $(0,1)$.
Whereas in the whole region I and at the point $(0,1)$, $\left\vert
F(t)\right\vert $ remains unity during the time evolution, and thus the QPT in
the environmental spin chain has no any effect on the entanglement of the two
central spins. This full localization behavior of $\left\vert F(t)\right\vert
$ can be seen from the analytic expression, Eq. (\ref{23}), in which $\tau
$=$0$ for $\gamma$=$0$, indicating no decay in $\left\vert F(t)\right\vert $,
regardless of the variation of $\lambda$ and the couping strength $g$.
Physically, this vanishing of disentanglement for the two central qubits under
an $XX$ spin-chain environment can be seen by noticing that the parameters
$\theta_{k}^{(\lambda_{j})}$ and $\theta_{k}^{(\lambda)}$ in Eq. (\ref{10})
are zero (or $\pi$) at $\gamma$=$0$. In this case, the fermionic modes
$b_{k,\lambda_{j}}$ and $b_{k,\lambda}$ coincide each other, which leads to
complete overlap between the ground state $|G\rangle_{\lambda_{j}}$ of
$H_{E}^{(\lambda_{j})}$ and the ground state $|G\rangle_{\lambda_{j^{\prime}}%
}$ of $H_{E}^{(\lambda_{j^{\prime}})}$, $|G\rangle_{\lambda_{j}}$%
=$|G\rangle_{\lambda_{j^{\prime}}}$. As a result, one sees from Eq. (\ref{14})
that the disentanglement factor $F(t)$ keeps an invariant value of unity
during its time evolution. Thus one arrives an important conclusion that the
enhancement of the disentanglement by QPT may be broken by special choice of
the spin-chain the occurrence of ground-state \textquotedblleft
accidental\textquotedblright\ degeneracy among the system-dressed
environmental spin-chain Hamiltonians $H_{E}^{(\lambda_{j})}$ in critical
parameter space. For further illustration, we show in Fig. 2 $\left\vert
F(t)\right\vert $ as a function of time and $\gamma$ for $\lambda$=1.0,
$N$=201, and $g$=0.05 (weak coupling), which corresponds to critical region
II. One can see that with deviating $\gamma$ from zero, the disentanglement
factor gradually evolves towards zero in an oscillatory way.%

\begin{figure}[tbp]
\begin{center}
\includegraphics[width=0.6\linewidth]{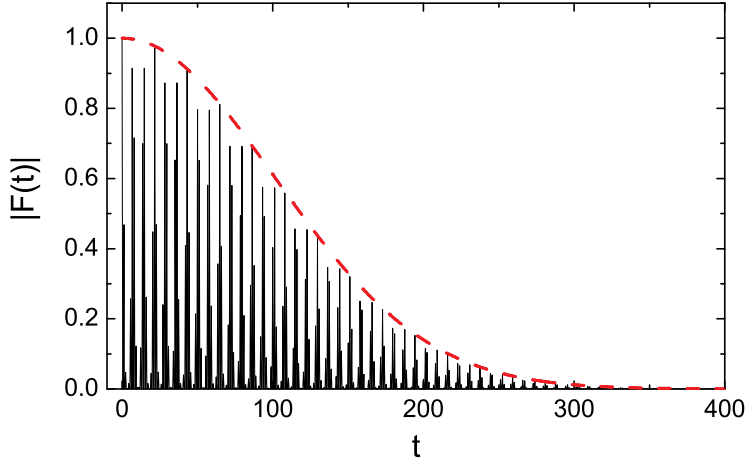}
\end{center}
\caption
{(Color online). Disentanglement factor $|F(t)|$ as a function of time in strong
coupling regime. The system parameters are chosen to be $\lambda=1.0$,
$N=201$, $\gamma=1.0$, and $g=500.0$. The exact numerical result is shown by
solid line, while the approximate Gaussian envelope factor is plotted by
dashed line.}
\label{fig3}
\end{figure}%
After discussing the QPT effect on the disentanglement of two spin qubits in
weak coupling regime ($g<<\lambda_{c}$), we turn now to study the QPT effect
in strong coupling regime ($g>>\lambda_{c}$). In Fig. 3 (solid line) we
display the time evolution of $|F(t)|$ for the values of $\lambda$=1.0,
$\gamma$=1.$0$ (Ising model), $N$=$201$, and $g$=$500$. Besides the
best-enhancement behavior ($|F(t)|\rightarrow$0 in final time) of the
disentanglement as discussed above, one additional prominent new feature,
which is absent in the weak coupling case, is that the decay of $|F(t)|$ is
now characterized by an oscillatory Gaussian envelope. To explain this, we
starts from the observation that when $g\gg1$, the spin-chain energy spectrum
in Eq. (\ref{8}) can be simplified to $\Omega_{k}^{(\lambda_{1})}%
\approx2\epsilon_{k}^{(\lambda_{1})}$ and $\Omega_{k}^{(\lambda_{2})}%
\approx-2\epsilon_{k}^{(\lambda_{2})}$. Thus from Eq. (\ref{9}) one has
$\theta_{k}^{(\lambda_{1})}\approx0$ and $\theta_{k}^{(\lambda_{2})}\approx
\pi$. This leads to the approximate identity $\alpha_{k}^{(\lambda_{1}%
)}-\alpha_{k}^{(\lambda_{2})}\approx-\pi/2$, by substitution of which into Eq.
(\ref{17.5})\ one can obtain
\begin{equation}
\left\vert F\left(  t\right)  \right\vert \approx\prod\nolimits_{k>0}|\cos
^{2}\alpha_{k}^{(\lambda_{1})}\exp\left(  i\bar{\Omega}_{k}t\right)  +\sin
^{2}\alpha_{k}^{(\lambda_{1})}\exp\left(  -i\bar{\Omega}_{k}t\right)
|,\label{e26}%
\end{equation}
where $\bar{\Omega}_{k}=\Omega_{k}^{\lambda_{1}}+\Omega_{k}^{\lambda_{2}}$.
Remarkably, the above expression for $\left\vert F\left(  t\right)
\right\vert $ is completely analogous to the one found when studying
decoherence on a qubit induced by noninteracting spin environment
\cite{Cucchietti2005} (see Eq. (16) in Ref. \cite{Cucchietti2005}). Thus, one
can exactly follow the mathematical derivation given in Ref.
\cite{Cucchietti2005} and Ref. \cite{Cucchietti2007}. The resultant
approximate expression for $\left\vert F\left(  t\right)  \right\vert $ is as
follows
\begin{equation}
\left\vert F\left(  t\right)  \right\vert \approx\exp\left(  -s_{N}^{2}%
t^{2}/2\right)  \left\vert \cos(\Omega t)\right\vert ^{(N-1)/2},\label{27}%
\end{equation}
where $\Omega$ is the mean value of $\bar{\Omega}_{k}$, i.e., $\Omega$%
=$\frac{1}{M}\sum_{k>0}\bar{\Omega}_{k}$, and
\begin{equation}
s_{N}^{2}=\sum_{k>0}\sin^{2}2\alpha_{k}^{(\lambda_{1})}\delta_{k}%
^{2}.\label{e28}%
\end{equation}
Here the quantity $\delta_{k}$ describes the deviation of $\bar{\Omega}_{k}$
from its mean value $\Omega$. It is straightforward to obtain $\Omega
\approx4g+\gamma^{2}/g$ and $\delta_{k}\approx-\frac{\gamma^{2}}{g}\cos
\frac{4\pi k}{N}$. We remark that the present Gaussian character of the
disentanglement factor is not only confined to the QPT regime. Here it is
mainly for purpose of the consistency in organizing this paper that we focus
our attention to the strongly coupling behavior of $\left\vert F\left(
t\right)  \right\vert $ in the vicinity of QPT. After a careful analysis of
Eq. (\ref{e28}), we further find that the width of the Gaussian envelope is
proportional to $g\gamma^{-2}N^{-1/2}$, which is an important scaling relation
between the decaying factor $\left\vert F\left(  t\right)  \right\vert $ and
the system parameters in the strong coupling QPT regime. For comparison with
the exact numerical result, we also show in Fig. 3 (dashed line) the Gaussian
envelope factor $\exp\left(  -s_{N}^{2}t^{2}/2\right)  $ in the approximate
expression (\ref{27}) of $\left\vert F\left(  t\right)  \right\vert $.
Clearly, the agreement is very good, indicating the validity of our
approximation in QPT region ($\lambda$=$1$ for Ising model) with strong
system-environment interaction.
\begin{figure}[tbp]
\begin{center}
\includegraphics[width=0.6\linewidth]{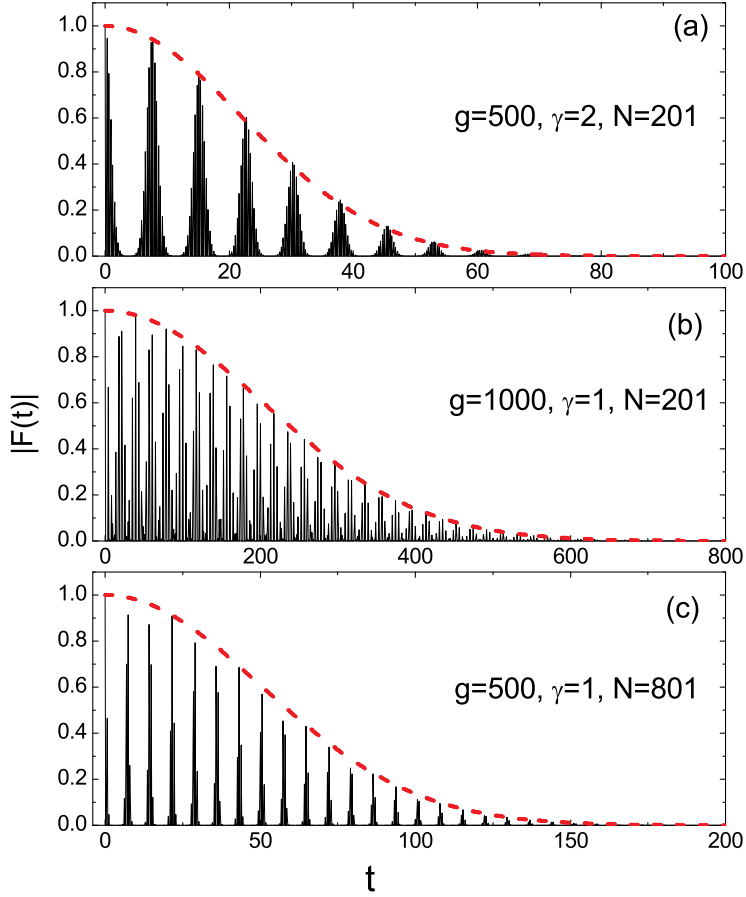}
\end{center}
\caption
{(Color online). Disentanglement factor $|F(t)|$ as a function of time in strong
coupling QPT regime ($\lambda
=1.0$) for different choices of parameters $\gamma
$, $g$ and $N$ to show their relationships with the decaying width of $|F(t)|$.
Again, the exact numerical result is shown by solid line, while the
approximate Gaussian envelope factor is plotted by dashed line.}
\label{fig4}
\end{figure}%
Figures 4(a)-(c) display the exactly numerical results (solid lines) of
$\left\vert F\left(  t\right)  \right\vert $ and the analytic results (dashed
lines) of Gaussian envelope factor $\exp\left(  -s_{N}^{2}t^{2}/2\right)  $
for $\lambda$=$1$ and different choices of the other parameters $g$, $\gamma$,
and $N$. It remarkably reveals in Fig. 4 that the decaying width of $|F(t)|$
is proportional to the product $g\gamma^{-2}N^{-1/2}$, exactly as we have
analyzed in the above discussions.%

\begin{figure}[tbp]
\begin{center}
\includegraphics[width=0.6\linewidth]{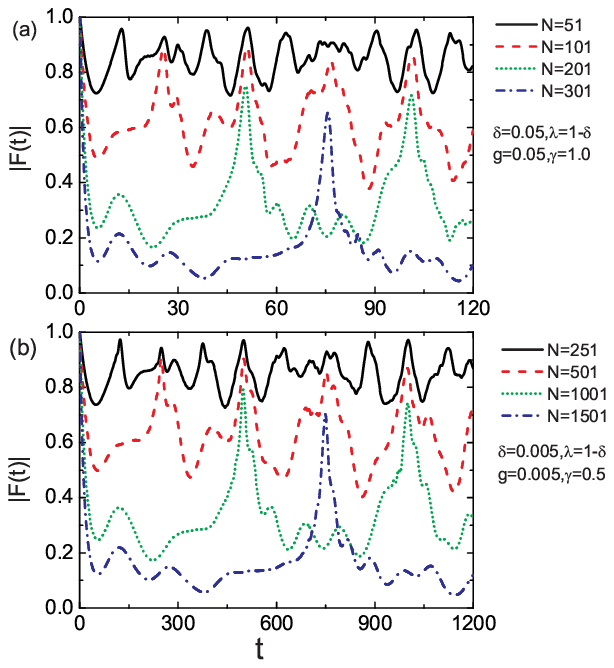}
\end{center}
\caption{(Color online). Scaling behavior of $|F(t)|$ in the vicinity of the
critical point $\lambda_{c}$=$1$ in the weak coupling case. The parameters
used in plotting curves in (b) are related to those used in plotting curves
(with the same curve type) in (a) by the transformation $g$ $\rightarrow\alpha
g$,
$\delta\rightarrow\alpha\delta$, $\gamma/N\rightarrow\alpha\gamma/N$ with
$\alpha=0.1$. One can see that by further transformation $t\rightarrow
t/\alpha$, figures (a) and (b) will completely overlap.}
\label{fig5}
\end{figure}%
Finally, we find that in the vicinity of QPT, the shape of the disentanglement
factor $\left\vert F\left(  t\right)  \right\vert $ during its time evolution
is invariant under the scaling transformation $t\rightarrow t/\alpha$,
$\delta\rightarrow\alpha\delta$, $g\rightarrow\alpha g$, and $\gamma
/N\rightarrow\alpha\gamma/N$, where $\delta=\lambda_{c}-\lambda$ characterizes
the vicinity of QPT. To illustrate this remarkable scaling property, we plot
in Figs. 5 the exact numerical results of evolution of $|F(t)|$ for different
values of the system parameters. Here the values of the system parameters used
in Fig. 5(b) are obtained from those used in Fig. 5(a) by a scaling factor
$\alpha=0.1$. Clearly, it shows in Fig. 5 that the exact time evolution of
$\left\vert F\left(  t\right)  \right\vert $ faithfully follow this scaling
transformation. Remarkably, the similar scaling property has been recently
found \cite{Quan} in studying dynamics of the LE for a single qubit coupled to
an Ising-type spin chain. Clearly, this scaling rule in the disentanglement
factor $|F(t)|$ for two entangled qubits or in the LE for the single qubit is
highly meaningful in quantum computing and quantum information processing.

To understand this scaling property, here we give a detailed analysis of the
behavior of $|F(t)|$ in the vicinity of the critical point $\lambda_{c}$=$1$
in the case of weak coupling strength $g$. Note that although in the present
context we only concern the specific model employed in this paper, the
following analysis can be easily applied to the other cases. We first notice
that in the expression of $|F(t)|$ [Eq. (\ref{18})], most factors $F_{k}$
remains nearly unity. Thus only very few $F_{k}$'s have remarkable effect on
the shape and amplitude of $|F(t)|$. From Eq (\ref{18}) one can see that in
order for the factor $F_{k}$ to deviate prominently from unity, at least one
of its two coefficients $\sin2\alpha_{k}^{(\lambda_{j})}$ ($j=1,2$) should be
considerably non-zero. Next let us check the value of $\sin2\alpha
_{k}^{(\lambda_{j})}$. For this we define $k_{c}^{(\lambda_{j})}$ which
enables $\left\vert \epsilon_{k_{c}}^{(\lambda_{j})}\right\vert $=$\left\vert
\lambda_{j}-\cos\left(  2\pi k_{c}^{(\lambda_{j})}/N\right)  \right\vert $ as
small as possible. For small $\delta$ (i.e., $\lambda-\lambda_{c}$) and $g$,
one can see that $k_{c}^{(\lambda_{j})}\ll M$. From the definition of
$\alpha_{k}^{(\lambda_{j})}$, we can write down
\begin{equation}
\sin2\alpha_{k}^{(\lambda_{j})}=\mp4\gamma g\sin\left(  2\pi k/N\right)
/\Omega_{k}^{(\lambda_{j})}\Omega_{k}^{\lambda}\label{33}%
\end{equation}
for $j=1$, 2, respectively. One can see from Eq. (\ref{33}) and the
expressions of $\Omega_{k}^{(\lambda_{j})}$ and $\Omega_{k}^{\lambda}$ that
for small $g$, to enable $\sin2\alpha_{k}^{(\lambda_{j})}$ considerably
non-zero, three conditions should be satisfied: (i) $k$ should be close to
$k_{c}^{(\lambda_{j})}$ and $k_{c}^{(\lambda)}$ in order for the amplitude of
$\gamma\sin\left(  2\pi k/N\right)  $ to be comparable with $\epsilon_{k_{c}%
}^{(\lambda_{j})}$ and $\epsilon_{k_{c}}^{(\lambda)}$; (ii) $g$ is small
enough so that $k$ could be close to $k_{c}^{(\lambda_{j})}$ and
$k_{c}^{(\lambda)}$ at the same time. (iii) $\delta$ is small which leads to
small value of $\sin\left(  2\pi k/N\right)  $ when $k$ approaching
$k_{c}^{(\lambda_{j})}$ and $k_{c}^{(\lambda)}$. Under these three conditions,
one has the following approximate expressions
\begin{align}
\Omega_{k}^{(\lambda_{j})} &  \approx2\left[  \left(  \delta\mp g\right)
^{2}+4\gamma^{2}\pi^{2}k^{2}/N^{2}\right]  ^{1/2},\label{34}\\
\Omega_{k}^{\lambda} &  \approx2\left(  \delta^{2}+4\gamma^{2}\pi^{2}%
k^{2}/N^{2}\right)  ^{1/2}.\nonumber
\end{align}
Combining Eq. (\ref{33}) and Eq. (\ref{34}), one immediately finds that the
transformation $g$ $\rightarrow\alpha g$, $\delta\rightarrow\alpha\delta$, and
$\gamma/N\rightarrow\alpha\gamma/N$ leads to $\Omega_{k}^{(\lambda_{j}%
)}\rightarrow\alpha\Omega_{k}^{(\lambda_{j})}$, $\Omega_{k}^{\lambda
}\rightarrow\alpha\Omega_{k}^{\lambda}$, while $\sin2\alpha_{k}^{(\lambda
_{j})}$ and $\cos2\alpha_{k}^{(\lambda_{j})}$ remaining invariant. As a
result, the time evolution of $|F(t)|$ in Eq. (\ref{18}) is well invariant
under further transformation $t\rightarrow t/\alpha$. This is what one has
seen from the exact results in Fig. 5.

In summary, we have studied the dynamic process of the disentanglement of a
coupled system consisting of two spin qubits and a general XY spin chain. The
exact expression of the disentanglement factor $|F(t)|$ has been obtained. The
relation between $|F(t)|$ and the QPT in the environmental spin chain has been
extensively illustrated. It has been shown that in general, the
disentanglement of the two qubits is best enhanced when the environmental spin
chain is exposed to QPT in either strong or weak coupling case. Both the
heuristic analysis and numerical calculations have shown the sharply decaying
behavior of the decoherence factor in the vicinity of the critical line
$\lambda$=$\lambda_{c}$=$1$. This decaying behavior, on the other side, has
been found to break for the particular XX spin chain ($\gamma=0$), in which
case $|F(t)|$ is not influenced by the environment. In the strong coupling
case, it has been numerically and analytically found that in the vicinity of
QPT the disentanglement factor decays to zero in an oscillatory Gaussian
envelope. The width of the Gaussian envelope has been found to scale with a
form $g\gamma^{-2}N^{-1/2}$. Furthermore, we have established a scaling rule
for the time evolution of the disentanglement factor in the vicinity of QPT.
We expect that the present results may shed light on the role of strongly
correlated environment played in the disentanglement dynamics of multi-qubits.

ZY and SL were supported by NSFC under Grant No. 60325416 and 60521001. PZ was
supported by NSFC under Grant Nos. 10604010 and 10544004.

\end{document}